\documentclass[preprint,aps,12pt,notitlepage,nofootinbib,tightenlines]{revtex4}
\usepackage{amsmath}
\usepackage{amssymb}
\usepackage{enumerate}
\usepackage{bm}
\usepackage{times}
\usepackage{braket}
\usepackage{color}
\usepackage{epsfig}
\usepackage{slashed}
\usepackage{hyperref}
\usepackage{multirow}
\usepackage{caption}
\usepackage{booktabs}
\usepackage{array}
\usepackage{float}
\usepackage[utf8]{inputenc}
\usepackage[T1]{fontenc}
\newcommand{\beq}{\begin{eqnarray}}
\newcommand{\eeq}{\end{eqnarray}}
\newcommand{\be}{\begin{equation}\begin{aligned}}
\newcommand{\ee}{\end{aligned}\end{equation}}

\newcommand{\gev}{\text{GeV}}

\definecolor{Red}{rgb}{1.,0.,0.}

\definecolor{Blue}{rgb}{0.,0.,1.}

\definecolor{nicered}{rgb}{0.7,0.1,0.1}
\definecolor{nicegreen}{rgb}{0.1,0.5,0.1}
\def\lsim{ {\ \lower-1.2pt\vbox{\hbox{\rlap{$<$}\lower6pt\vbox{\hbox{$\sim$}}}}\ } }
\def\gsim{ {\ \lower-1.2pt\vbox{\hbox{\rlap{$>$}\lower6pt\vbox{\hbox{$\sim$}}}}\ } }

\hypersetup{colorlinks,citecolor=nicegreen,linkcolor=nicered}
\begin{document}
\title{Searches for heavy neutrinos at a 3 TeV CLIC in fat jet final states}
\author{Yao-Bei Liu$^{1,2}$\footnote{E-mail: liuyaobei@hist.edu.cn}, Jing-Wei Lian$^{1,3}$, }
\affiliation{1. Henan Institute of Science and Technology, Xinxiang 453003, China\\
2. School of Electro-Mechanical Engineering, Zhongyuan Institute of Science and Technology, Xuchang 461000, China\\
3. Department of Physics, Henan Normal University, Xinxiang 453007, China}
\begin{abstract}
Heavy Majorana neutrinos ($N$) are predicted  in many models of physics beyond
the Standard Model. In this work, we explore the production and detection prospects of TeV-scale heavy neutrinos ($m_N \gtrsim 1$ TeV) at a future 3 TeV Compact Linear Collider (CLIC). We focus on two distinct decay topologies: (i) $N \to \ell^\pm W^\mp$ with hadronic $W$ boson decay, leading to a final state with one charged lepton and a hadronic fat-jet $J_W$; and (ii) $N \to \nu h$ with subsequent Higgs decay $h \to b\bar{b}$, yielding a Higgs-tagged fat-jet $J_h$ and $\slashed{E}_T$. Based on comprehensive detector-level simulations and background analysis, we present both $2\sigma$ exclusion limits and $5\sigma$ discovery reaches in the $m_N$–$|V_{\ell N}|^2$ plane. We further extract 95\% confidence level upper limits on the  mixing parameter $|V_{\ell N}|^2$, and perform a detailed comparison with existing constraints from direct searches at future colliders and indirect global fits. Our findings demonstrate that a 3 TeV CLIC can improve the sensitivity to $|V_{\ell N}|^2$ by about two orders of magnitude compared to the projected reaches of future hadron colliders, while remaining competitive with other CLIC search channels.
\end{abstract}

\maketitle

\newpage
\section{Introduction}
Heavy Majorana neutrinos ($N$) provide critical insights into the origin of neutrino masses and lepton number violation (LNV)~\cite{Minkowski:1977sc,Yanagida:1979as,Gell-Mann:1979vob,Glashow:1979nm,Mohapatra:1979ia}, where the type-I seesaw framework introduces gauge-singlet right-handed neutrinos $(N_{R})$. Through their mixing with Standard Model~(SM) left-handed neutrinos $\nu_L$, heavy mass eigenstates $N$ containing small $\nu_L$ components emerge, with masses ranging from eV to $10^{14}$ GeV depending on model parameters. These scenarios are further constrained by neutrinoless double beta decay ($0\nu\beta\beta$) experiments that probe effective Majorana masses at the sub-eV scale~\cite{Agostini:2022zub}. Consequently, experimental searches for LNV processes involving heavy neutrinos play a vital role in establishing the Majorana nature of neutrinos~\cite{Atre:2009rg,Cai:2017mow}.

The production cross section and  decay width of the heavy neutrino $N$ are critically dependent on both its mass $m_N$ and the mixing parameter $|V_{\ell N}|^2$, which characterizes the flavor-dependent mixing between $N$ and the SM neutrino $\nu_\ell$. Current Large Hadron Collder~(LHC) searches have constrained $|V_{\ell N}|^2$ through investigations of LNV signatures, particularly in same-sign dilepton plus dijet final states~\cite{CMS:2018iaf,ATLAS:2019kpx,CMS:2019lwf,CMS:2018jxx,ATLAS:2018dcj,CMS:2022hvh,CMS:2024bni,CMS:2024xdq}. For neutrino masses between 10 and 50\,GeV, the most stringent limits on $|V_{\ell N}|^2$ reach approximately $10^{-5}$~\cite{CMS:2018iaf,ATLAS:2019kpx}, although these constraints weaken significantly above the $Z$ boson mass threshold.
Very recently, a stringent upper limit on the mixing parameter $|V_{\ell N}|^2$ in the range $8.0\times 10^{-4}$--$6.3\times 10^{-2}$ was established for heavy neutrino masses $m_N = 90$--600~GeV. This constraint was derived from a combined analysis of the $\mu^{\pm}\mu^{\mp} e^{\pm}\nu$ and $\mu^{\pm}\mu^{\mp} \mu^{\pm}\nu$ channels at the 13~TeV LHC with an integrated luminosity of 138~fb$^{-1}$~\cite{delaTorre:2025nki}. Complementary bounds on the mixing parameters for masses below 50\,GeV can be derived from $W$ boson decay analyses~\cite{LHCb:2020wxx}.  Notably, displaced vertex searches have achieved even stronger sensitivity, reaching limits of order $10^{-6}$~\cite{CMS:2022fut}, albeit limited to the low-mass regime below 10--15\,GeV.

While TeV-scale heavy neutrino production at the LHC suffers from suppressed cross sections, limiting constraints on active-sterile mixing~\cite{delAguila:2008hw,Pascoli:2018heg,Dev:2013wba,Helo:2013esa,Das:2015toa,Antusch:2018bgr,Babu:2022ycv,Bolton:2024thn}, alternative colliders offer complementary probes: electron-proton facilities enable unique signatures~\cite{Liang:2010gm,Blaksley:2011ey,Duarte:2014zea,Mondal:2016kof,Antusch:2016ejd,Lindner:2016lxq,Li:2018wut,Das:2018usr,Antusch:2019eiz,Das:2020gnt,Gu:2022muc,Gu:2022nlj,Yang:2023ice,Shen:2025}, and muon colliders access extreme mass ranges~\cite{Mekala:2023diu,Kwok:2023dck,Li:2023tbx,Yang:2023ojm,Jiang:2023mte,Li:2023lkl,Wang:2023zhh,He:2024dwh,Cao:2024rzb},
Future \( e^+e^- \) colliders will provide a unique opportunity to probe heavy neutrinos~\cite{Fargion:1995qb}. The International Linear Collider (ILC)~\cite{ILC:2013jhg,ILCInternationalDevelopmentTeam:2022izu} has been extensively studied for heavy neutrino masses \( m_N \lesssim 500 \) GeV, with sensitivity explored in various channels~\cite{Zhang:2018rtr,Lu:2022wsm,Wang:2016eln,delAguila:2005pin,delAguila:2005ssc,Das:2012ze,Banerjee:2015gca,Hernandez:2018cgc,Biswal:2017nfl,Bellagamba:2025xpd}. At higher masses (\( m_N \sim \mathcal{O}(1) \) TeV and beyond), the Compact Linear Collider (CLIC)~\cite{CLICDetector:2013tfe,Franceschini:2019zsg}, with planned center-of-mass energies of \( \sqrt{s} = 1.4 \) TeV and 3 TeV, will offer enhanced discovery potential. Recent studies highlight the sensitivity of the CLIC across multiple signatures: Ref.~\cite{Chakraborty:2018khw} reports \( 5\sigma \) reaching \( |V_{eN}|^2 \sim 10^{-5} \)--\( 10^{-6} \) at 500 fb\(^{-1}\) via \( N \to e^\pm W^\mp \) for \( m_N = 600 \)--2700 GeV, whereas Ref.~\cite{Mekala:2022cmm} extends the coverage to \( qq\ell \) final states (Dirac/Majorana) for \( M_N = 200 \) GeV--3.2 TeV. Additionally, the \( e^-\gamma \) mode achieves \( \mathcal{O}(10^{-5}) \) sensitivity at \( 2\sigma \) for same-sign dileptons at \( \sqrt{s} = 3 \) TeV with \( m_N = 1 \)--2.5 TeV~\cite{Das:2023tna}.

In this work, we investigate the production of heavy Majorana neutrinos ($m_N = 1000$--2900\,GeV) at a 3\,TeV CLIC, assuming flavor-symmetric mixings $|V_{eN}|^2 = |V_{\mu N}|^2$. We focus on the dominant decay channels $N \to \ell^\pm W^\mp$ and $N \to \nu h$, where the high center-of-mass energy of the CLIC produces highly boosted final states. This enables efficient identification of hadronic decays ($W \to q\bar{q'}$, $h \to b\bar{b}$) through their characteristic fat-jet signatures ($J_W$, $J_h$). Our analysis establishes both $2\sigma$ exclusion limits and $5\sigma$ discovery potential, demonstrating the unique sensitivity of the CLIC to previously unexplored regions of the heavy neutrino parameter space beyond current experimental reach.

This remainder of this paper is organized as follows: Section~\ref{sec:model} presents the theoretical framework for Majorana neutrino production and decay. Section~\ref{sec:analysis} details the collider analysis methodology, including signal and background simulations. Finally, we present a summary in Section~\ref{sec:conclusion}.

\section{Model\label{sec:model}}
Heavy neutrinos, as outlined in the introduction, are predicted by various seesaw models~\cite{Mohapatra:1979ia,Magg:1980ut,Cheng:1980qt,Foot:1988aq}. While the minimal Type-I seesaw mechanism typically predicts heavily suppressed mixing angles for TeV-scale heavy neutrinos, this prediction is in direct tension with the experimental sensitivities explored in this work. However, several well-motivated extensions, such as the inverse seesaw~\cite{Mohapatra:1986aw,Mohapatra:1986bd,Nandi:1985uh} or other models with protected symmetry, can naturally yield large mixing angles. Therefore, for this study on the sensitivity reach of future high-energy $e^{+}e^{-}$ colliders, we adopt a phenomenological and model-independent approach, considering only mixing-induced interactions between the new neutrinos and the SM. Thus, our analysis is performed within the $SM\_HeavyN\_LO$ framework for Majorana neutrinos, which effectively extends the SM by introducing three right-handed neutrino singlets ($N_1$, $N_2$, $N_3$) under the SM gauge groups~\cite{delAguila:2008cj,Alva:2014gxa,Degrande:2016aje}.

The model's Lagrangian consists of the SM terms supplemented by new interactions:
\begin{equation}
\mathcal{L} = \mathcal{L}_{SM} + \mathcal{L}_{N} + \mathcal{L}_{WN\ell} + \mathcal{L}_{ZN\nu} + \mathcal{L}_{HN\nu}
\end{equation}
where $\mathcal{L}_{N}$ contains the kinetic and mass terms for the heavy neutrinos (expressed in 4-spinor notation throughout):
\begin{equation}
\mathcal{L}_{N} = \frac{1}{2}\sum_{k=1}^3 \left(\bar{N}_{k}i\slashed{\partial}N_{k} - m_{N_k}\bar{N_{k}}^{C}N_{k}\right),
\end{equation}
with the sum explicitly indicating the three neutrino states, and the superscript C denotes the charge conjugation.

The interaction Lagrangians for heavy neutrinos with SM gauge and Higgs bosons are given by
\begin{align}
\mathcal{L}_{WN\ell} &= - \frac{g}{\sqrt{2}}W^{+}_{\mu} \sum^{3}_{k=1}\sum^{\tau}_{\ell=e}\bar{N}_{k}V^{*}_{\ell k}\gamma^{\mu}P_{L}\ell^{-} + \text{h.c.}, \\
\mathcal{L}_{ZN\nu} &= - \frac{g}{2\cos\theta_{W}}Z_{\mu} \sum^{3}_{k=1}\sum^{\tau}_{\ell=e}\bar{N}_{k}V^{*}_{\ell k}\gamma^{\mu}P_{L}\nu_{\ell} + \text{h.c.}, \\
\mathcal{L}_{HN\nu} &= - \frac{gm_{N}}{2M_{W}}h \sum^{3}_{k=1}\sum^{\tau}_{\ell=e}\bar{N}_{k}V^{*}_{\ell k}P_{L}\nu_{\ell} + \text{h.c.},
\end{align}
where $V_{\ell k}$ represents the heavy-light neutral lepton mixing matrix elements.

The partial decay widths for the three dominant heavy neutrino decay channels are
\begin{align}
\Gamma(N \rightarrow \ell W) &= \frac{g^2}{64\pi}|V_{\ell N}|^{2}\frac{m_N^3}{M_W^2}\left(1-\frac{M_W^2}{m_N^2}\right)^2\left(1+2\frac{M_W^2}{m_N^2}\right), \\
\Gamma(N \rightarrow \nu_\ell Z) &= \frac{g^2}{128\pi}|V_{\ell N}|^{2}\frac{m_N^3}{M_W^2}\left(1-\frac{M_Z^2}{m_N^2}\right)^2\left(1+2\frac{M_Z^2}{m_N^2}\right), \\
\Gamma(N \rightarrow \nu_\ell h) &= \frac{g^2}{128\pi}|V_{\ell N}|^2\frac{m_N^3}{M_W^{2}}\left(1-\frac{M_h^{2}}{m_N^{2}}\right)^2.
\label{eq:partial_widths}
\end{align}

In the heavy mass limit ($m_N \gg M_W,M_Z,M_h$), the branching ratios approach the simple ratio
\begin{equation}
\text{BR}(N\rightarrow \ell W) : \text{BR}(N\rightarrow \nu Z) : \text{BR}(N\rightarrow \nu h) \simeq 2 : 1 : 1.
\end{equation}
For our region of interest ($m_N \geq 1000$ GeV), this corresponds to $\text{BR}(N\rightarrow \ell W) \approx 50\%$, making the charged lepton channel particularly prominent for heavy neutrino searches.

\section{Collider simulation and analysis\label{sec:analysis}}
\subsection{Production cross section}
To simplify our analysis, we consider a minimal scenario with a single generation of heavy Majorana neutrino $N$ that mixes exclusively with electron and muon flavor active neutrinos. This framework assumes flavor-symmetric mixing parameters $|V_{\ell N}|^2 = |V_{e N}|^2 = |V_{\mu N}|^2 \neq 0$ (for $\ell = e, \mu$), while setting $|V_{\tau N}|^2 = 0$. The remaining heavy neutrinos $N_2$ and $N_3$ are decoupled with masses fixed at 10 TeV and all their mixing parameters set to zero. For reference sample generation, the mixing parameter $|V_{\ell N}|^2$ has been set to $10^{-4}$.  For our choice of parameter space, the heavy neutrino has a microscopic
 lifetime ($c\tau\ll 1$ nm) and no displaced vertices are expected.

\begin{figure}[thb]
\begin{center}
\vspace{-0.5cm}
\centerline{\epsfxsize=10cm \epsffile{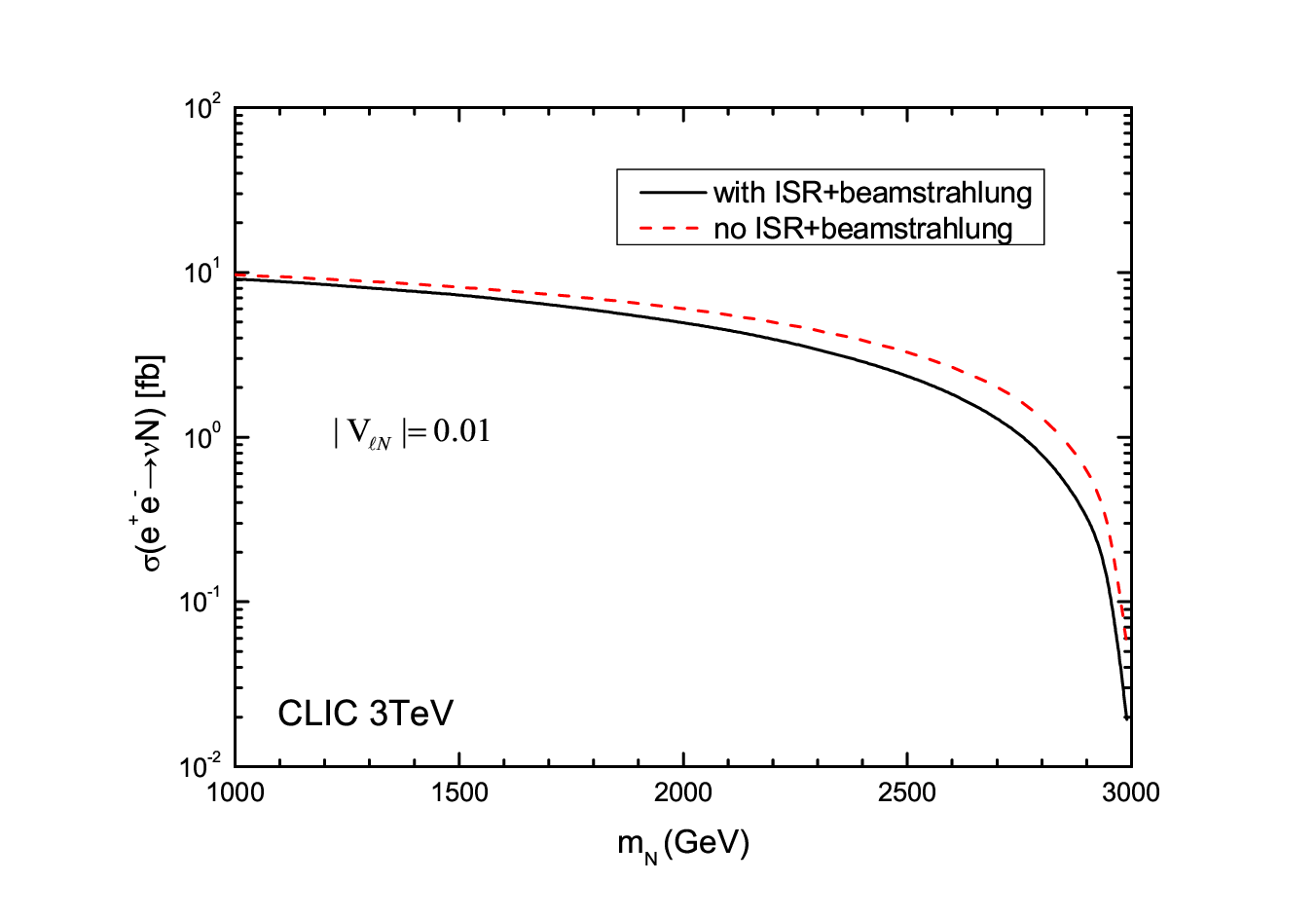}}
\caption{The production cross section of $e^{+}e^{-}\to \nu N$ as a function of $m_N$ at the 3 TeV CLIC with $|V_{\ell N}|^2 = |V_{e N}|^2 = |V_{\mu N}|^2=10^{-4}$.}
\label{cross}
\end{center}
\end{figure}

At $e^+e^-$ colliders, the dominant production mechanisms proceed through $s$-channel $Z$ boson and $t$-channel $W$ boson exchanges.
Figure~\ref{cross} shows the calculated cross sections $\sigma(e^+e^- \to \nu N)$ as a function of $m_N$ for a fixed mixing parameter $|V_{\ell N}|^2 = 10^{-4}$  at a center-of-mass energy of $\sqrt{s}=3~$TeV, comparing scenarios with and without initial state radiation (ISR) and beamstrahlung effects. These calculations, performed at leading order using \textsc{MadGraph5\_aMC@NLO}~\cite{mg5}, reveal that ISR effects moderately reduce the production cross section across the mass range. For instance, at $m_N = 2$ TeV, the cross section decreases from 6.03 fb (without ISR) to 4.95 fb (with ISR), whereas the overall production rate naturally decreases with increasing $m_N$ owing to phase space suppression.

\begin{figure}[h]
\centering
\includegraphics[width = 16cm ]{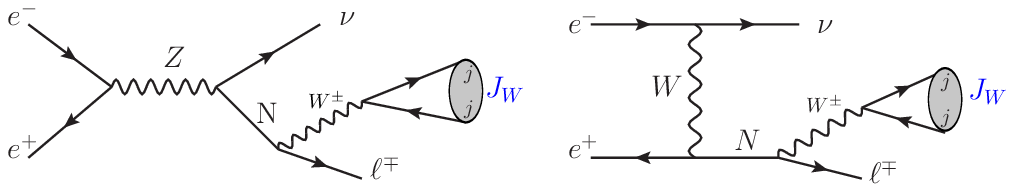}\\
\vspace{-18cm}
\includegraphics[width = 16cm ]{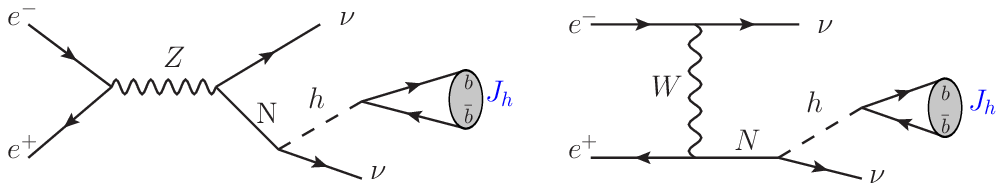}
\vspace{-18cm}
\caption{Feynman diagrams of processes for Cases 1 and 2.}
\label{fey}
\end{figure}

We investigate the discovery potential of heavy Majorana neutrinos $N$ through comprehensive Monte Carlo simulations of signal and background processes at CLIC energies. Our analysis focuses on two characteristic decay channels: $N \to \ell^{\pm}W^{\mp}$ and $N \to \nu h$, where the high boost of the $W$ and Higgs bosons leads to collimated hadronic decays forming distinctive fat-jets ($J_W$ and $J_h$ respectively). The key final-state signatures are characterized by
\begin{align*}
\text{Case 1:} \quad & e^{+}e^{-} \to \nu N \to \ell^{\pm}W^{\mp}\nu \to \ell^{\pm} + J_{W} + \slashed{E}_T, \\
\text{Case 2:} \quad & e^{+}e^{-} \to \nu N \to \nu \nu h \to J_{h} + \slashed{E}_T,
\end{align*}

\noindent where $J_W$ represents a 2-prong fat jet from $W\to q\bar{q'}$ decay, $J_h$ is a $b$-tagged 2-prong jet from $h\to b\bar{b}$, $\slashed{E}_T$ denotes missing transverse energy from all neutrino contributions, and $\ell^{\pm}$ refers to isolated $e^{\pm}$ or $\mu^{\pm}$ leptons.
The complete production and decay topology is illustrated in Fig.~\ref{fey}, showing both the $t$-channel $W$-exchange and $s$-channel $Z$-exchange processes.

We simulate the signal and background processes using a complete Monte Carlo workflow. The event generation is performed at leading order using \textsc{MadGraph5\_aMC@NLO}~\cite{mg5}, with subsequent parton showering and hadronization handled by \textsc{Pythia} 8.20~\cite{pythia8}. Detector effects are incorporated through fast simulation with \textsc{Delphes} 3.4.2~\cite{deFavereau:2013fsa}, using the dedicated CLIC detector card (\texttt{delphes\_card\_CLICdet\_Stage3\_fcal.tcl})~\cite{Leogrande:2019qbe} that accounts for the unique detector geometry and performance specifications.
For jet reconstruction, we employ the Valencia Linear Collider (VLC) algorithm~\cite{Boronat:2014hva,Boronat:2016tgd} with parameters ($R = 1.0$, $\beta = 1$, $\gamma = 1$) optimized for the high-energy lepton collider environment. The $b$-jet identification uses the medium working point with 70\% efficiency for $b$-quark jets, along with corresponding misidentification rates for charm and light jets. The final event analysis and statistical interpretation are conducted using \textsc{MadAnalysis5}~\cite{Conte:2012fm,Conte:2014zja}.

\begin{table}[ht!]
\centering
\caption{Cross sections (in fb) of signals for two cases at 3 TeV CLIC with $|V_{\ell N}|^2=10^{-4}$, including ISR and beamstrahlung effects.\label{cs:signal}}
\renewcommand{\arraystretch}{0.9}  
\footnotesize
\begin{tabular}{@{}cccc|cccc@{}}
\hline
\textbf{Mass~(GeV)} & \textbf{Case 1} & \textbf{Case 2} & & \textbf{Mass~(GeV)} & \textbf{Case 1} & \textbf{Case 2} \\[0.5ex]  
\hline
1000 & 2.85 & 1.24 & & 2000 & 1.54 & 0.68 \\[-0.5ex]  
1100 & 2.75 & 1.19 & & 2100 & 1.39 & 0.62 \\[-0.5ex]
1200 & 2.63 & 1.15 & & 2200 & 1.22 & 0.55 \\[-0.5ex]
1300 & 2.52 & 1.11 & & 2300 & 1.06 & 0.47 \\[-0.5ex]
1400 & 2.38 & 1.05 & & 2400 & 0.90 & 0.40 \\[-0.5ex]
1500 & 2.27 & 1.00 & & 2500 & 0.73 & 0.33 \\[-0.5ex]
1600 & 2.12 & 0.94 & & 2600 & 0.57 & 0.25 \\[-0.5ex]
1700 & 1.99 & 0.88 & & 2700 & 0.41 & 0.18 \\[-0.5ex]
1800 & 1.84 & 0.82 & & 2800 & 0.26 & 0.12 \\[-0.5ex]
1900 & 1.70 & 0.75 & & 2900 & 0.11 & 0.05 \\[-0.5ex]
\hline
\end{tabular}
\end{table}

\begin{table}[ht!]
\centering
\caption{SM background processes and their corresponding cross sections at the 3 TeV CLIC, including ISR and beamstrahlung effects.\label{cs:sm}}
\renewcommand{\arraystretch}{1.1}
\begin{tabular}{@{}lcp{4.5cm}c@{}}
\hline
\textbf{Case} & \textbf{SM Process} & \textbf{Decay Channels} & \textbf{Cross section (fb)} \\[0.5ex]
\hline
\multirow{2}{*}{Case 1} & $e^{+}e^{-} \to \ell^{\pm}\nu jj$ & -- & 499 \\
                         & $e^{+}e^{-} \to \nu\bar{\nu}W^{+}W^{-}$ & $W \to \ell\nu$, $W \to q\bar{q}'$ & 33.8 \\[1ex]  \hline
\multirow{3}{*}{Case 2} & $e^{+}e^{-} \to \nu\bar{\nu}h$ & $h \to b\bar{b}$ & 269 \\
                         & $e^{+}e^{-} \to \nu\bar{\nu}Z$ & $Z \to b\bar{b}$ or $Z \to q\bar{q}$ & 1410 \\
                         & $e^{+}e^{-} \to W^{+}W^{-}$ & $W \to \ell\nu$, $W \to q\bar{q}'$ & 115 \\
\hline
\end{tabular}
\end{table}

 The dominant SM backgrounds are categorized by final state topology: for Case 1 ($\ell^{\pm} + J_W + \slashed{E}_T$), the main backgrounds result from $e^{+}e^{-}\to\ell^{\pm}\nu jj$ processes (comprising on-shell $WW$ production, $t$-channel $W$-exchange diagrams, and off-shell gauge boson contributions) as well as $e^{+}e^{-}\to\nu\bar{\nu}W^{+}W^{-}$ with mixed decays ($W\to\ell\nu$ plus $W \to q \bar{q'}$) and fully hadronic $W$ decays; for Case 2 ($J_h + \slashed{E}_T$), the dominant backgrounds include $e^{+}e^{-}\to\nu\bar{\nu}h$ with $h\to b\bar{b}$, $e^{+}e^{-}\to\nu\bar{\nu}Z$ with the $Z$ boson decaying hadronically, and $e^{+}e^{-}\to W^{+}W^{-}$ with mixed decays where one $W$ decays hadronically, whereas the other decays leptonically with the charged lepton $\ell$ escaping detection, and processes with negligible cross sections after selection cuts (such as $Zhh$, $ZZh$, and $t\bar{t}h$) are omitted from our analysis.

  Tables~\ref{cs:signal} and~\ref{cs:sm} summarize the cross sections for both signal processes and SM backgrounds at the $\sqrt{s} = 3$ TeV CLIC, including the effects of the ISR and beamstrahlung. The signal cross sections are calculated assuming $|V_{\ell N}|^2 = 10^{-4}$ for two benchmark scenarios, wheras the background processes include all relevant production channels with their dominant decay modes.

To identify objects, we select the basic cuts at parton level for the signals and SM backgrounds as follows:
 \be
p_{T}^{\ell/j}>~10~\gev,\quad
 |\eta_{\ell}|<~2.5,\quad
 |\eta_{j}|<~5\\
  \ee
where $p_{T}^{\ell, j}$ and $\eta_{\ell/j}$ are the transverse momentum  and pseudorapidity, respectively, of leptons and jets.

\subsection{Case 1: $1\ell+J_{W}+\slashed{E}_{T}$ analysis}
\begin{figure*}[htb]
\begin{center}
\centerline{\hspace{2.0cm}\epsfxsize=9cm\epsffile{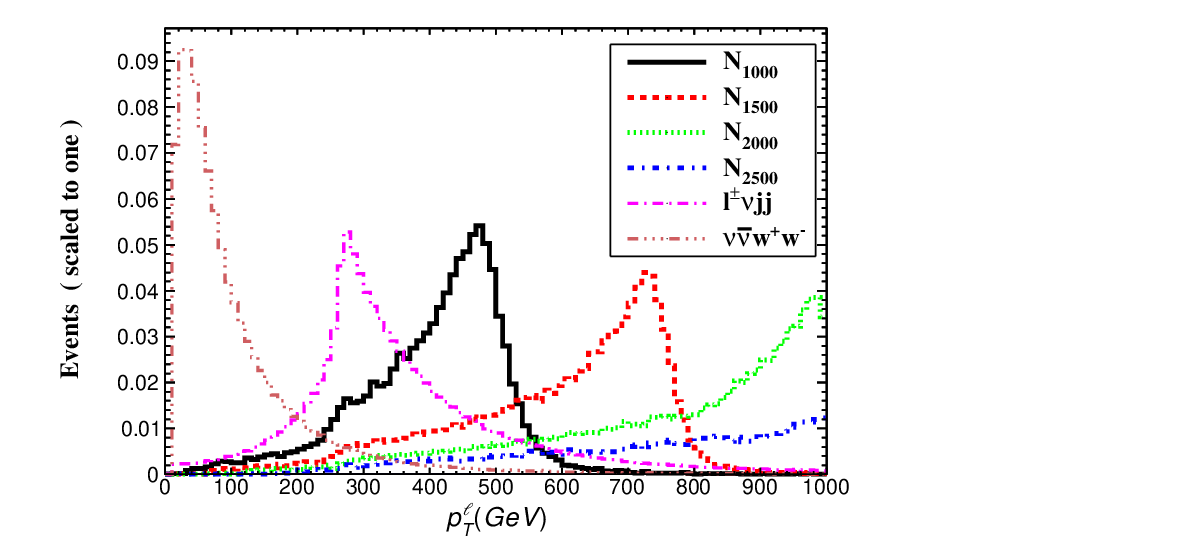}
\hspace{-2.0cm}\epsfxsize=9cm\epsffile{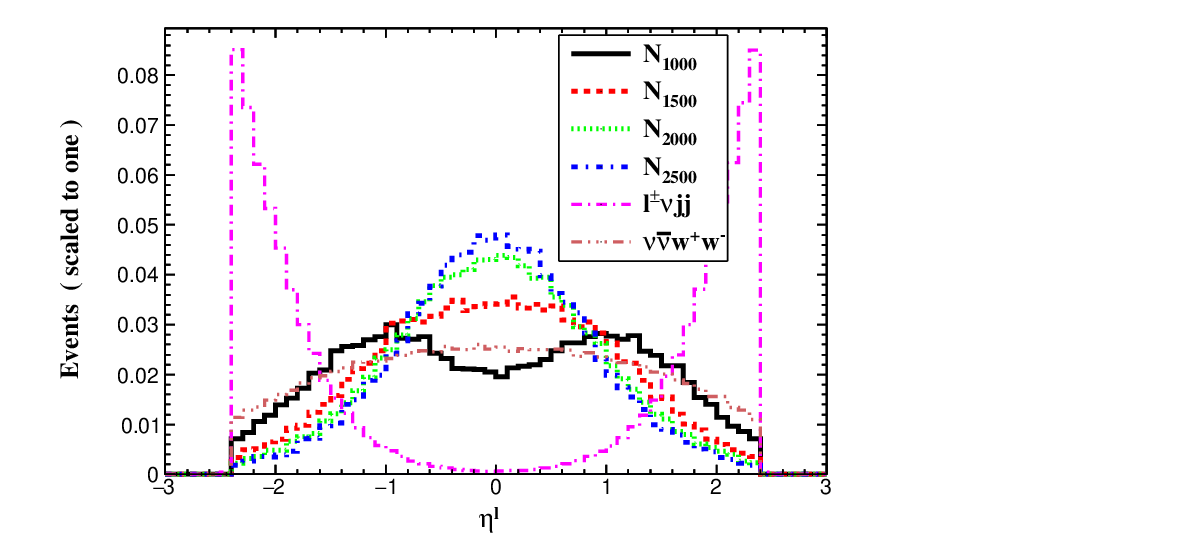}}
\centerline{\hspace{2.0cm}\epsfxsize=9cm\epsffile{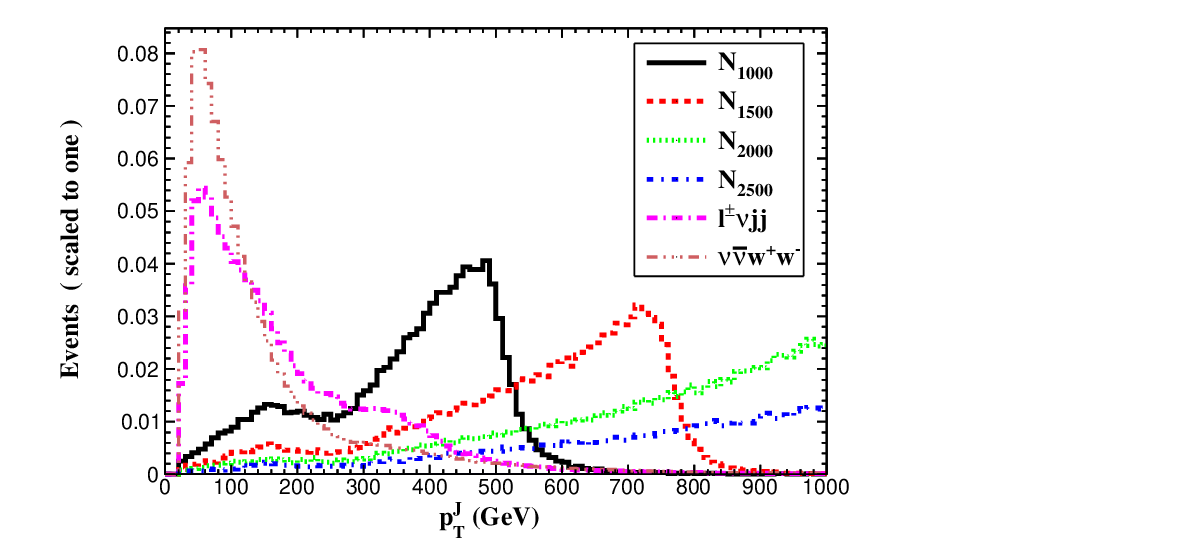}
\hspace{-2.0cm}\epsfxsize=9cm\epsffile{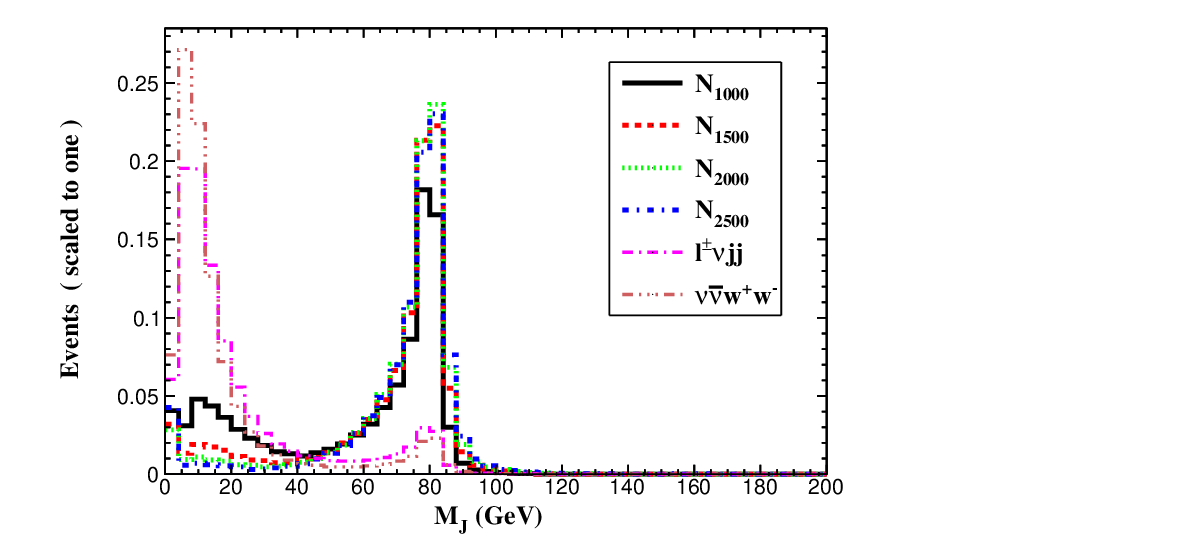}}
\centerline{\hspace{2.0cm}\epsfxsize=9cm\epsffile{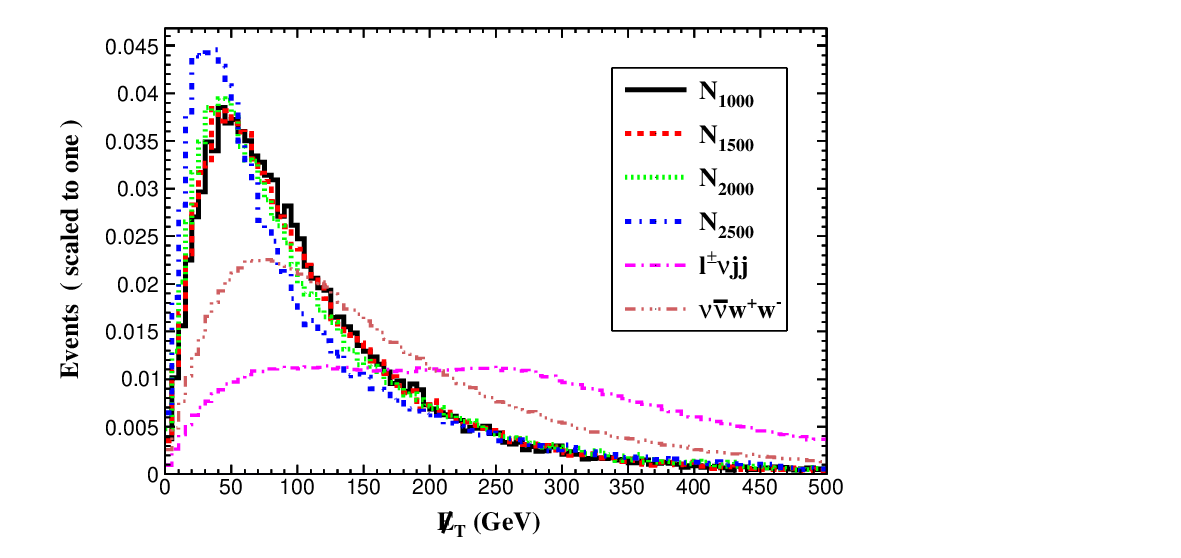}
\hspace{-2.0cm}\epsfxsize=9cm\epsffile{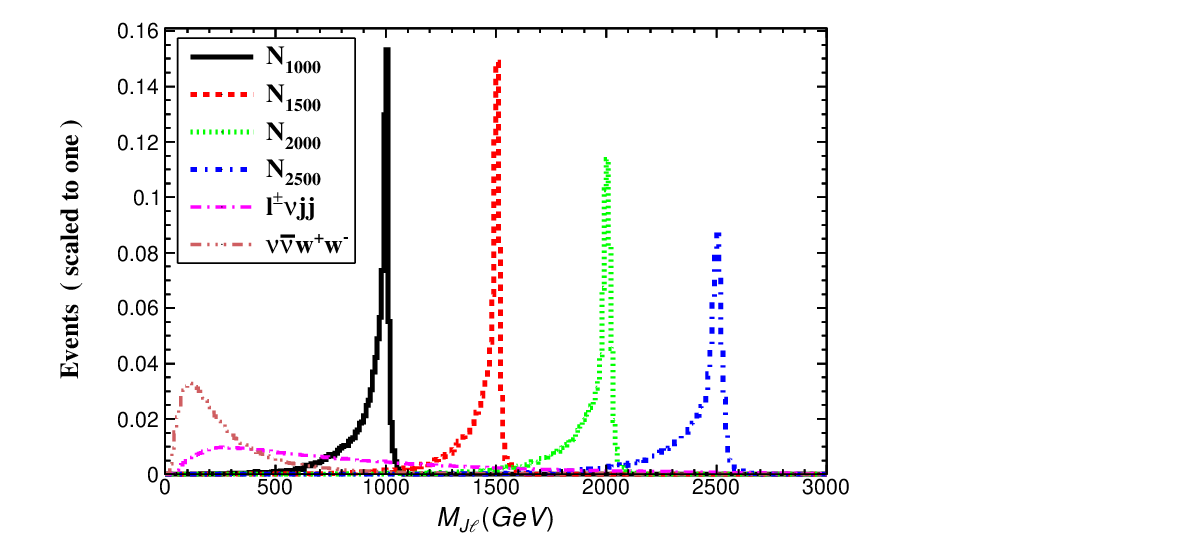}}
\caption{Normalized distributions for the four signals (with $m_N = 1000$ , 1500, 2000,  and 2500 GeV and relevant SM backgrounds for Case 1. }
\label{fig:distribution1}
\end{center}
\end{figure*}
Figure~\ref{fig:distribution1} displays the differential distributions for signal events with Majorana neutrino masses $m_N = 1000$, 1500, 2000, and 2500~GeV, along with relevant SM backgrounds. The distributions include the lepton and fat jet transverse momenta ($p_T^{\ell,J}$), lepton pseudorapidity ($\eta^\ell$), fat jet invariant mass ($M_J$), missing transverse energy ($\slashed{E}_T$), and  reconstructed neutrino mass ($M_{\ell J}$).
The heavy Majorana neutrino leads to distinctive kinematic features: its decay products are highly boosted, producing leptons and fat jets with significantly larger transverse momentum compared to background processes. Signal leptons predominantly appear in the central region ($\eta^\ell \approx 0$), whereas background leptons tend to populate the forward region. Background processes dominate the high $\slashed{E}_T$ region owing to contributions from multiple neutrinos and potential detector effects. The invariant mass $M_{\ell J}$ distribution clearly reveals the mass peaks for different $m_N$ hypotheses.

We employ the following sequential selection criteria to optimize signal extraction:

\begin{itemize}
    \item \textbf{Cut-1: Missing energy cut}: $\slashed{E}_T < 150$~GeV

    \item \textbf{Cut-2: Fat jet selection}:
    \begin{itemize}
        \item $p_T^J > 300$~GeV
        \item $|M_J - m_W| < 15$~GeV (consistent with hadronic $W$ decay)
    \end{itemize}

    \item \textbf{Cut-3: Lepton selection}:
    \begin{itemize}
        \item Exactly one charged lepton ($N(\ell)=1$)
        \item $|\eta^\ell| < 1.5$
        \item Transverse momentum requirements:
        \begin{itemize}
            \item $p_T^\ell > 350$~GeV for $1000 \leq m_N \leq 1400$~GeV
            \item $p_T^\ell > 500$~GeV for $1500 \leq m_N \leq 1900$~GeV
            \item $p_T^\ell > 600$~GeV for $m_N \geq 2000$~GeV
        \end{itemize}
    \end{itemize}

    \item \textbf{Cut-4: Invariant mass cut}:
    \begin{itemize}
        \item $M_{\ell J} > 900$~GeV for $1000 \leq m_N \leq 1400$~GeV
        \item $M_{\ell J} > 1400$~GeV for $1500 \leq m_N \leq 1900$~GeV
        \item $M_{\ell J} > 1900$~GeV for $m_N \geq 2000$~GeV
    \end{itemize}
\end{itemize}

\begin{table}[htb]
\centering
\caption{Cut flow efficiencies for signal benchmarks ($m_N = 1000$, 1500, and 2000~GeV) and dominant backgrounds in Case 1.\label{cutflow1}}
\vspace{0.2cm}
\begin{tabular}{lcccccc}
\toprule[1.5pt]
\multirow{2}{*}{Selection Cuts} & \multicolumn{3}{c}{Signal Efficiency} & & \multicolumn{2}{c}{Background Efficiency} \\
\cmidrule{2-4} \cmidrule{6-7}
 & 1000 GeV & 1500 GeV & 2000 GeV & & $\ell^{\pm}\nu jj$ & $\nu\bar{\nu}W^{+}W^{-}$ \\
\midrule[1pt]
$\sigma_{0}~(fb)$ & 2.85 & 2.27 & 1.54 & & 499 & 33.8 \\
\hline
Cut 1 ($\slashed{E}_T$) & 0.75 & 0.75 & 0.75 & & 0.28 & 0.54 \\[-0.2em]
Cut 2 (Fat jet) & 0.46 & 0.60 & 0.64 & & 0.04 & 0.004 \\[-0.2em]
Cut 3 (Lepton) & & & & & & \\[-0.5em]
\quad $m_N$=1000 GeV & 0.28 & -- & -- & & 0.003 & 0.0018 \\[-0.2em]
\quad $m_N$=1500 GeV & -- & 0.39 & -- & & 0.0021 & 0.0012 \\[-0.2em]
\quad $m_N$=2000 GeV & -- & -- & 0.44 & & 0.0015 & 0.00083 \\[-0.2em]
Cut 4 ($M_{\ell J}$) & & & & & & \\[-0.5em]
\quad $m_N$=1000 GeV & 0.27 & -- & -- & & 0.0024 & 0.0017 \\[-0.2em]
\quad $m_N$=1500 GeV & -- & 0.35 & -- & & 0.0017 & 0.001 \\[-0.2em]
\quad $m_N$=2000 GeV & -- & -- & 0.38 & & 0.0012 & 0.00069 \\
\bottomrule[1.5pt]
\end{tabular}
\end{table}

We report the efficiencies for signal processes at three benchmark mass points ($m_N = 1000$, 1500, and 2000 GeV) and their corresponding SM backgrounds following the sequential application of selection cuts, as detailed in Table~\ref{cutflow1}. The cross sections after each selection stage are computed using
$\sigma_{\text{after cut}} = \sigma_0 \times \epsilon_{\text{cut}}$,
where $\sigma_0$ represents the initial production cross section, and $\epsilon_{\text{cut}}$ denotes the cumulative efficiency up to that cut. The incremental efficiency for each individual cut can be determined from the ratio of consecutive cross-section values.

After the full selection criteria are applied, the total background cross section is reduced to $1.27~\text{fb}$ in the mass range $1000 \leq m_N \leq 1400~\text{GeV}$, decreasing further to $0.86~\text{fb}$ for $1500 \leq m_N \leq 1900~\text{GeV}$ and reaching $0.59~\text{fb}$ for the heavy mass regime ($m_N \geq 2000~\text{GeV}$). The signal cross sections after all cuts are presented in Table~\ref{cut:s1}.

\begin{table}[ht!]
\centering
\caption{Signal cross sections (in fb) after all selection cuts for Case~1 at 3 TeV CLIC with $|V_{\ell N}|^2=10^{-4}$.\label{cut:s1}}
\renewcommand{\arraystretch}{0.9}  
\footnotesize
\begin{tabular}{@{}ccc|ccc@{}}
\hline
\textbf{Mass~(GeV)} & \textbf{Cross section}  & & \textbf{Mass~(GeV)} & \textbf{Cross section}  \\[0.5ex]  
\hline
1000 & 0.77 & & 2000 & 0.58 \\[-0.5ex]  
1100 & 0.91 & & 2100 & 0.63 \\[-0.5ex]
1200 & 0.95 & & 2200 &0.56 \\[-0.5ex]
1300 & 0.99  & & 2300 & 0.51 \\[-0.5ex]
1400 & 0.98 & & 2400 & 0.43 \\[-0.5ex]
1500 & 0.79& & 2500 & 0.36 \\[-0.5ex]
1600 & 0.87 & & 2600 & 0.29 \\[-0.5ex]
1700 & 0.85 & & 2700 & 0.21 \\[-0.5ex]
1800 & 0.83& & 2800 & 0.13 \\[-0.5ex]
1900 & 0.76 & & 2900 & 0.06 \\[-0.5ex]
\hline
\end{tabular}
\end{table}

\subsection{Case 2: $J_{h}+\slashed{E}_{T}$ analysis }

\begin{figure*}[htb]
\begin{center}
\centerline{\hspace{2.0cm}\epsfxsize=9cm\epsffile{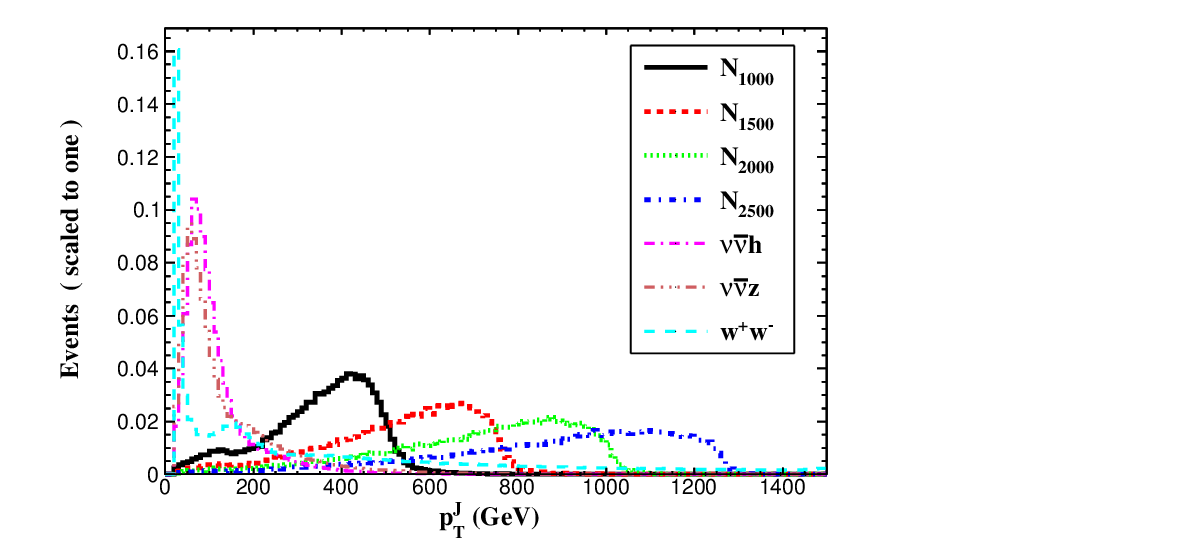}
\hspace{-2.0cm}\epsfxsize=9cm\epsffile{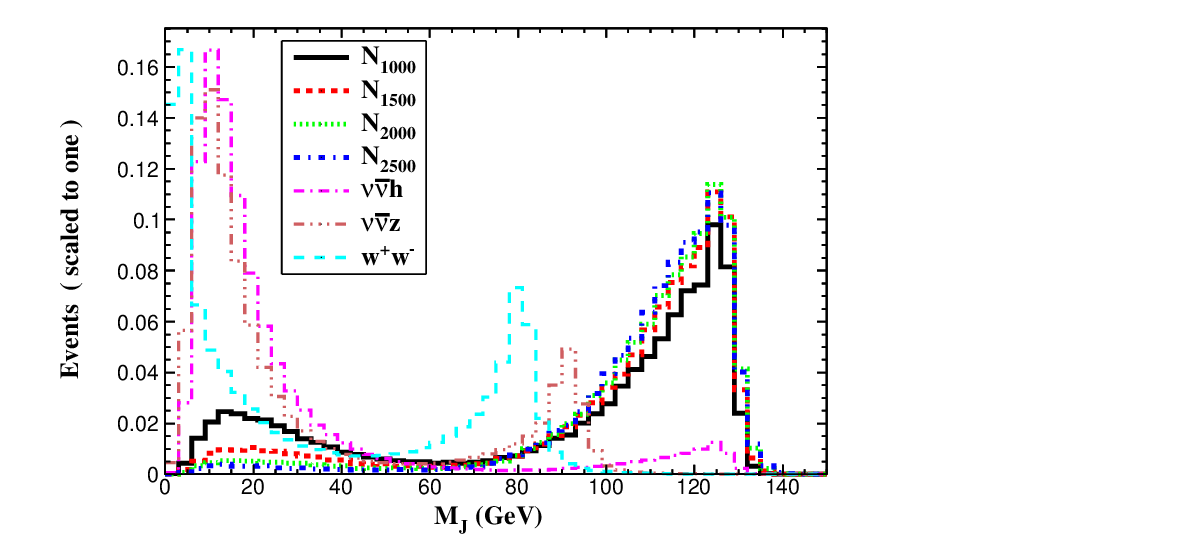}}
\centerline{\hspace{2.0cm}\epsfxsize=9cm\epsffile{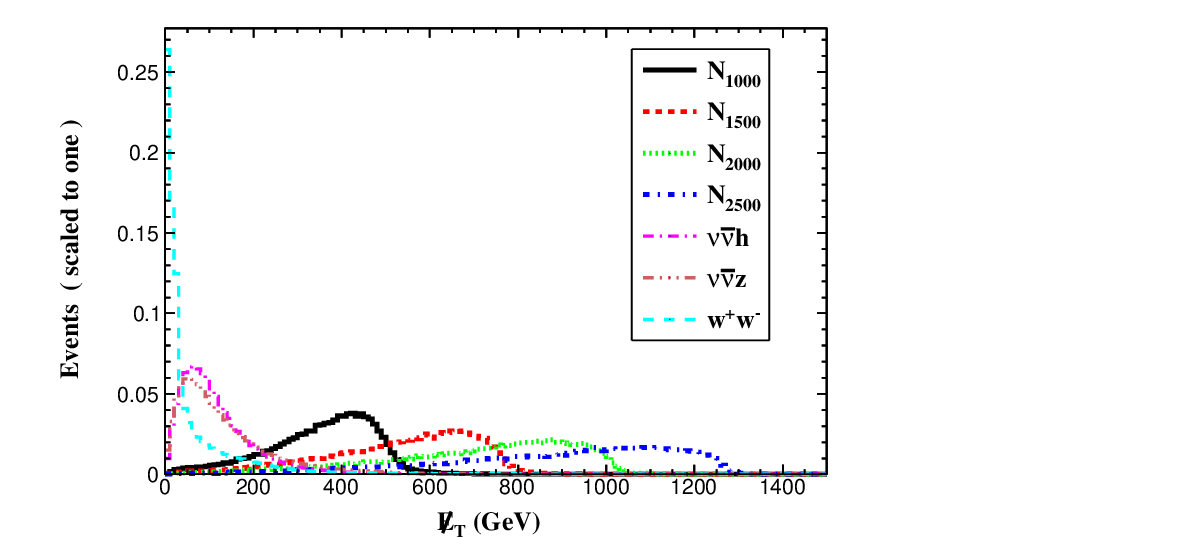}}
\caption{Normalized distributions for the signals and relevant SM backgrounds for Case 2. }
\label{fig:distribution2}
\end{center}
\end{figure*}

To optimize background suppression, we systematically analyze the normalized distributions of key kinematic variables for signal benchmarks ($m_N = 1000$, 1500, 2000, and 2500$~\text{GeV}$) and SM backgrounds (Fig.~\ref{fig:distribution2}). The studied variables include: fat jet transverse momentum ($p_T^J$), fat jet invariant mass ($M_J$),
and missing transverse energy ($\slashed{E}_T$).
The analysis reveals distinct kinematic features: signal events exhibit significantly higher missing transverse energy compared to SM backgrounds, wheras the invariant mass distribution of Higgs-originated fat jets ($J_h$) exhibits a clear peak around the Higgs boson mass.

Based on these kinematic studies, we implement the following optimized selection criteria:

\begin{itemize}
    \item \textbf{Cut-1: Lepton veto}: Reject events with $N(\ell) \geq 1$ to suppress leptonic backgrounds

    \item \textbf{Cut-2: Fat jet selection}:
    \begin{itemize}
        \item Invariant mass requirement: $M_J > 100~\text{GeV}$
        \item Transverse momentum thresholds (signal-mass-dependent):
        \begin{itemize}
            \item $p_T^J > 300~\text{GeV}$ for $1000 \leq m_N \leq 1400~\text{GeV}$
            \item $p_T^J > 400~\text{GeV}$ for $1500 \leq m_N \leq 1900~\text{GeV}$
            \item $p_T^J > 500~\text{GeV}$ for $m_N \geq 2000~\text{GeV}$
        \end{itemize}
    \end{itemize}

    \item \textbf{Cut-3: Missing energy requirements}:
    \begin{itemize}
        \item $\slashed{E}_T > 350~\text{GeV}$ for $1000 \leq m_N \leq 1400~\text{GeV}$
        \item $\slashed{E}_T > 550~\text{GeV}$ for $1500 \leq m_N \leq 1900~\text{GeV}$
        \item $\slashed{E}_T > 700~\text{GeV}$ for $m_N \geq 2000~\text{GeV}$
    \end{itemize}
\end{itemize}

\begin{table}[htb]
\centering
\caption{Cut flow efficiencies for signal benchmarks ($m_N = 1000$, 1500, and 2000~GeV) and dominant SM backgrounds in Case 2.\label{cutflow2}}
\begin{tabular}{lccccccc}
\toprule[1.5pt]
 & \multicolumn{3}{c}{Signal} & & \multicolumn{3}{c}{Background} \\
\cmidrule(lr){2-4} \cmidrule(lr){6-8}
Cut & 1000 GeV & 1500 GeV & 2000 GeV & & $\nu\bar{\nu}h$ & $\nu\bar{\nu}Z$ & $W^+W^-$ \\
\midrule[1pt]
$\sigma_0$ (fb) & 1.24 & 1.00 & 0.68 & & 269 & 1411 & 115 \\
\addlinespace[-2pt]
Cut 1 (Lepton veto) & 1.0 & 1.0 & 1.0 & & 1.0 & 1.0&0.77 \\
\addlinespace[-2pt]
Cut 2 (Fat jet) & & & & & & & \\
\quad $m_N$=1000 GeV & 0.55 & -- & -- & & 0.036 & 0.0038&6.07e-4 \\
\addlinespace[-2pt]
\quad $m_N$=1500 GeV & -- & 0.65 & -- & & 0.015 & 0.0021&6.05e-4 \\
\addlinespace[-2pt]
\quad $m_N$=2000 GeV & -- & -- & 0.69 & & 0.007 & 0.0012&5.74e-4 \\
\addlinespace[-2pt]
Cut 3 ($\slashed{E}_T$) & & & & & & &\\
\quad $m_N$=1000 GeV & 0.47 & -- & -- & & 0.023&2.84e-3 & 2.05e-4 \\
\addlinespace[-2pt]
\quad $m_N$=1500 GeV & -- & 0.48 & -- & & 4.77e-3 & 9.0e-4&1.28e-4 \\
\addlinespace[-2pt]
\quad $m_N$=2000 GeV & -- & -- & 0.53 & & 1.77e-3& 4.17e-4&9.2e-5 \\
\bottomrule[1.5pt]
\end{tabular}
\end{table}

We report the selection efficiencies for signal benchmarks at three representative mass points ($m_N = 1000$, 1500, and 2000 GeV) and their corresponding SM backgrounds following the sequential kinematic selections detailed in Table~\ref{cutflow2}. The analysis demonstrates significant background suppression exceeding two orders of magnitude while maintaining signal efficiencies above 45\% across all mass points.
The complete selection criteria reduce the total background cross section to $10.26~\text{fb}$ ($1000 \leq m_N \leq 1400~\text{GeV}$), $2.57~\text{fb}$ ($1500 \leq m_N \leq 1900~\text{GeV}$), and $1.07~\text{fb}$ ($m_N \geq 2000~\text{GeV}$).
The signal cross sections after all cuts are listed in Table~\ref{cut:s2}.

\begin{table}[ht!]
\centering
\caption{Signal cross sections (in fb)  after full selection cuts for Case~2 at 3 TeV CLIC with $|V_{\ell N}|^2=10^{-4}$.\label{cut:s2}}
\renewcommand{\arraystretch}{0.9}  
\footnotesize
\begin{tabular}{@{}ccc|ccc@{}}
\hline
\textbf{Mass~(GeV)} & \textbf{Cross section}  & & \textbf{Mass~(GeV)} & \textbf{Cross section}  \\[0.5ex]  
\hline
1000 & 0.58 & & 2000 & 0.36 \\[-0.5ex]  
1100 &0.65 & & 2100 & 0.35 \\[-0.5ex]
1200 & 0.69 & & 2200 &0.33 \\[-0.5ex]
1300 & 0.69  & & 2300 & 0.29 \\[-0.5ex]
1400 & 0.69 & & 2400 & 0.26 \\[-0.5ex]
1500 & 0.48& & 2500 & 0.22 \\[-0.5ex]
1600 & 0.51 & & 2600 & 0.17 \\[-0.5ex]
1700 & 0.51 & & 2700 & 0.12 \\[-0.5ex]
1800 & 0.49& & 2800 & 0.073 \\[-0.5ex]
1900 & 0.48 & & 2900 & 0.035 \\[-0.5ex]
\hline
\end{tabular}
\end{table}

\subsection{Statistical analysis}
The discovery ($\mathcal{Z}_{\text{disc}}$) and exclusion ($\mathcal{Z}_{\text{excl}}$) significances are computed using~\cite{Cowan:2010js}
\beq
 \mathcal{Z}_\text{disc} &=& \sqrt{2[(s+b)\ln(1+s/b)-s]}, \\
 \mathcal{Z}_\text{excl} &=& \sqrt{2[s-b\ln(1+s/b)]},
\eeq
 where $s$ and $b$ denote the expected number of signal and background events, respectively, for a given integrated luminosity~($\mathcal{L}_{int}$).

\begin{figure}[htb]
\begin{center}
\vspace{-0.5cm}
\centerline{\epsfxsize=9cm \epsffile{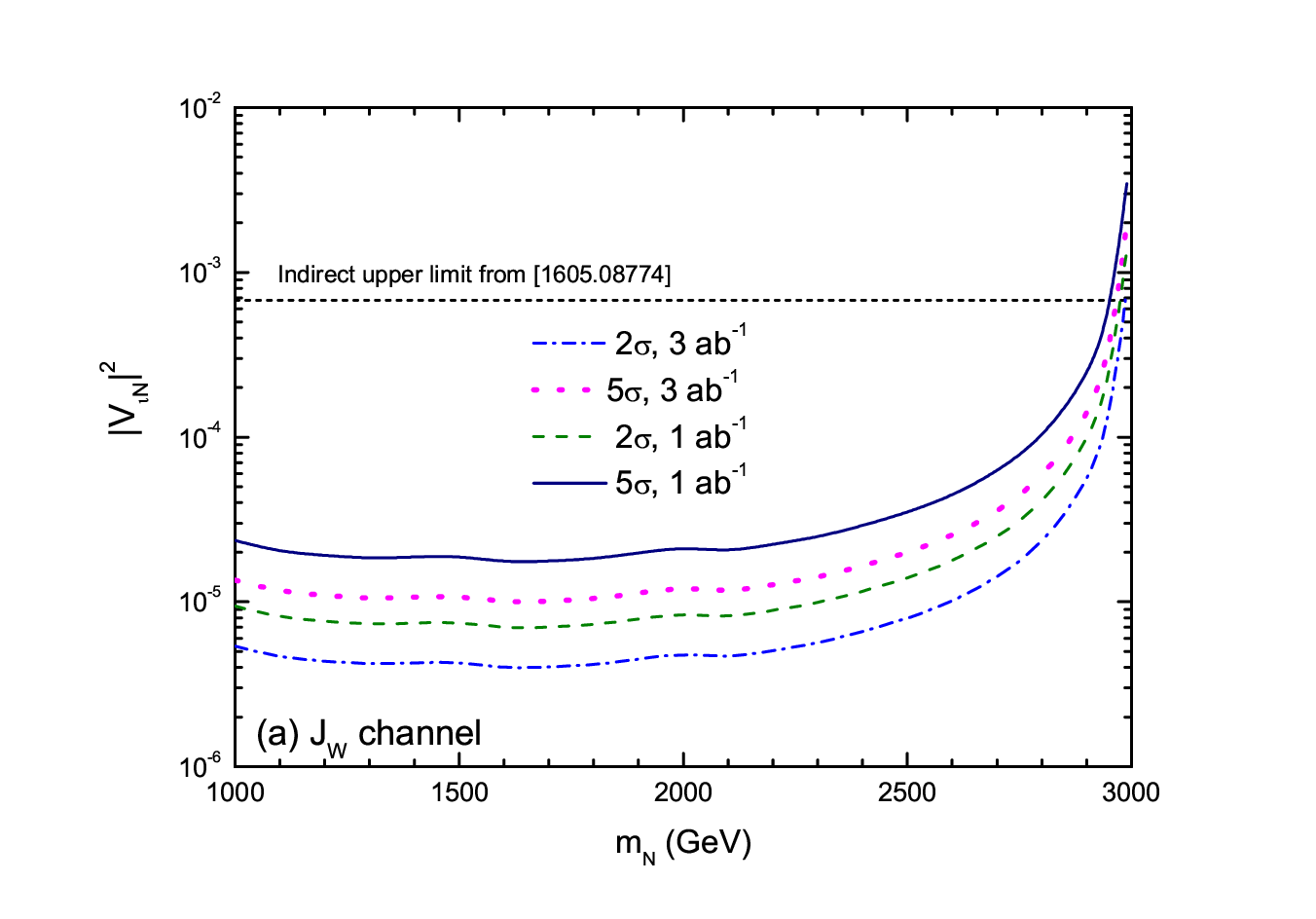}\epsfxsize=9cm \epsffile{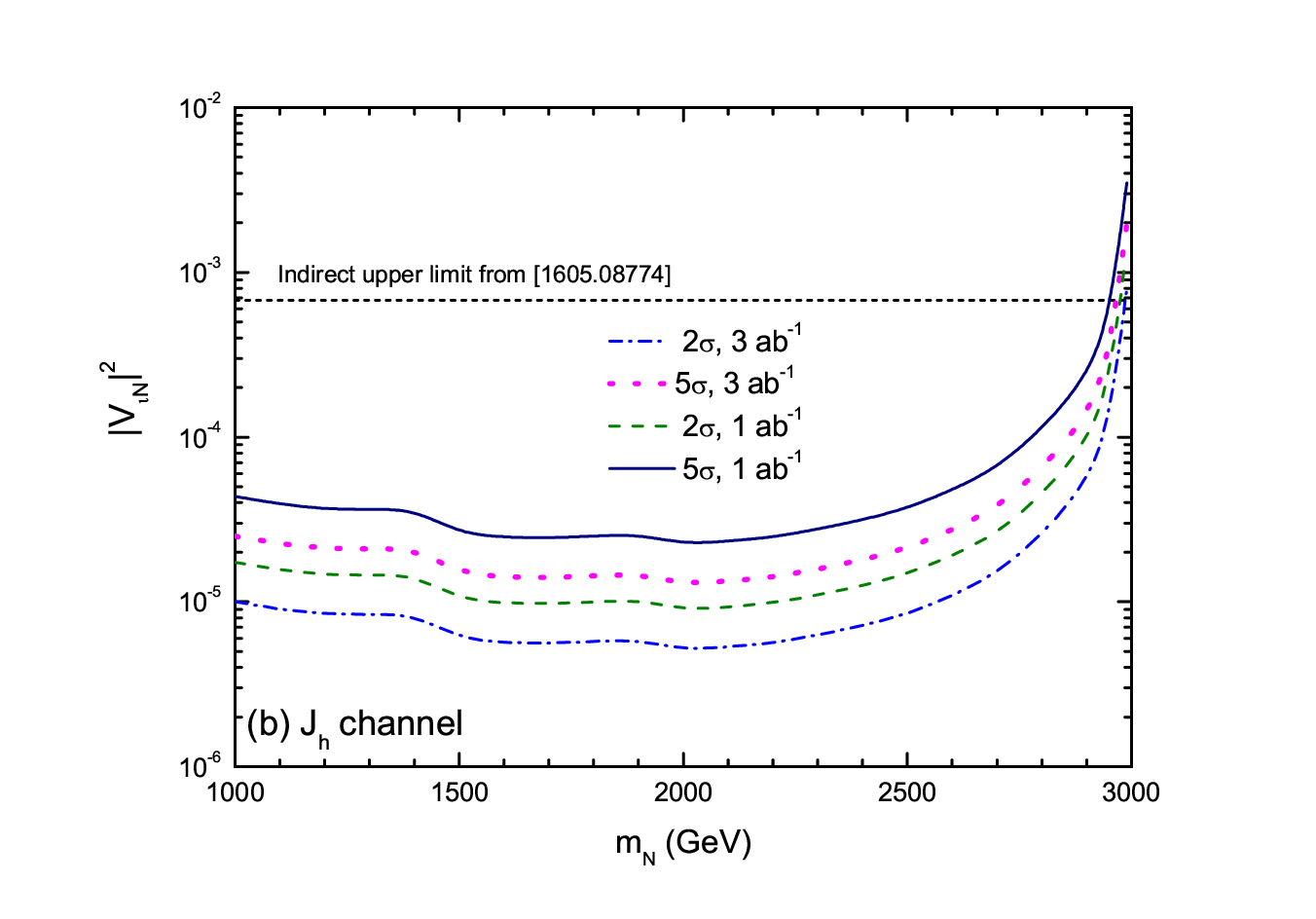}}
\caption{Exclusion limits ($2\sigma$) and discovery reach ($5\sigma$) for (a): Case 1, and (b): Case 2 in the $m_N$-$|V_{\ell N}|^2$ parameter space. }
\label{fig-ss}
\end{center}
\end{figure}
\begin{figure}[t]
\begin{center}
\vspace{1.5cm}
\centerline{\epsfxsize=12cm \epsffile{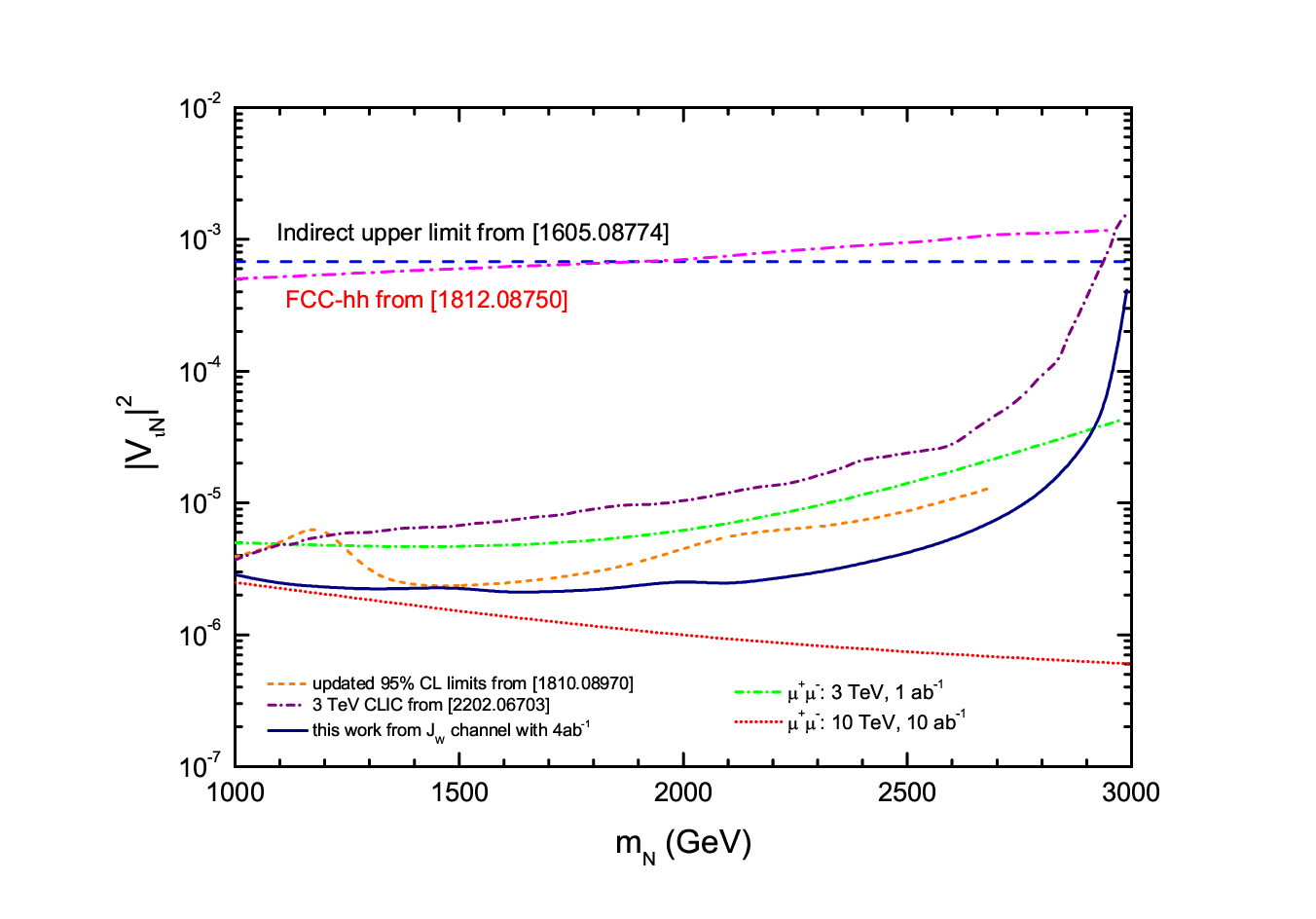}}
\caption{95\% CL exclusion limits in the $|V_{\ell N}|^{2}-m_N$ plane for 3~TeV CLIC ($\mathcal{L}_{int}=4~\text{ab}^{-1}$, $P_{e^+}=0$, $P_{e^-}=-0.8$), compared with: FCC-hh direct searches~\cite{Pascoli:2018heg}, $\mu^{+}\mu^{-}$ colliders~\cite{Li:2023tbx}, previous 3~TeV CLIC studies~\cite{Chakraborty:2018khw,Mekala:2022cmm} with identical luminosity and polarization,  and indirect global constraints~\cite{Fernandez-Martinez:2016lgt}.}
\label{fig6}
\end{center}
\end{figure}

Figure~\ref{fig-ss} presents the $2\sigma$ exclusion limits and $5\sigma$ discovery reaches in the $|V_{\ell N}|^2$--$m_N$ plane for a 3~TeV CLIC with $\mathcal{L}_{int} = 1\text{--}3~\text{ab}^{-1}$, assuming flavor-symmetric mixings $|V_{\ell N}|^2 = |V_{e N}|^2 = |V_{\mu N}|^2$. The CLIC sensitivity exceeds indirect global constraints~\cite{Fernandez-Martinez:2016lgt} for $m_N \lesssim 2900~\text{GeV}$. Across $m_N = 1000\text{--}2500~\text{GeV}$, $2\sigma$ limits of $\mathcal{O}(10^{-6}\text{--}10^{-5})$ are found for Case 1, which provides stronger constraints than Case 2. For $m_N = 2~\text{TeV}$, the $5\sigma$ discovery reach for Case 1 is $|V_{\ell N}|^2 = 2.1 \times 10^{-5}$ ($1~\text{ab}^{-1}$) and $1.2 \times 10^{-5}$ ($3~\text{ab}^{-1}$), compared with $5.1 \times 10^{-5}$ and $2.9 \times 10^{-5}$ for Case 2. Similarly, the $2\sigma$ exclusion limits reach $8.6 \times 10^{-6}$ ($1~\text{ab}^{-1}$) and $4.9 \times 10^{-6}$ ($3~\text{ab}^{-1}$) for Case 1, and $2.0 \times 10^{-5}$ and $1.1 \times 10^{-5}$ for Case 2.

Recent studies of heavy neutrinos at the 3~TeV CLIC have employed complementary approaches. Chakraborty et al.~\cite{Chakraborty:2018khw} studied flavor-specific mixing ($|V_{eN}|^2\neq0$) via the $e^+e^-\to\nu_e N\to e^\pm W^\mp$ channel (with hadronic $W$ decays) at both $\sqrt{s}=1.4$ and 3~TeV, although their 5$\sigma$ sensitivity projections used only 500~fb$^{-1}$ of unpolarized data. In contrast, Mekala et al.~\cite{Mekala:2022cmm} analyzed $qq\ell$ final states using Boosted Decision Trees (BDT) with 4~ab$^{-1}$ integrated luminosity and $-80\%$ electron polarization (unpolarized positrons), obtaining 95\% CL limits on $|V_{\ell N}|^2$ for masses $m_N\in[200\,\text{GeV},3.2\,\text{TeV}]$ under the universal mixing hypothesis $V^2_{eN} = V^2_{\mu N} = V^2_{\tau N}$.
Moreover, future muon colliders can probe $|V_{\mu N}|^2$ down to $\mathcal{O}(10^{-6}-10^{-5})$ at $\sqrt{s}=3$~TeV (1~ab$^{-1}$) and $\mathcal{O}(10^{-7}-10^{-6})$ at $\sqrt{s}=10$~TeV (10~ab$^{-1}$), respectively~\cite{Li:2023tbx}.

Figure~\ref{fig6} compares our results in Case 1 with these previous constraints and includes FCC-hh projections~\cite{Pascoli:2018heg} for $pp\to 2\mu \ell_X$ processes (30~fb$^{-1}$) as a hadron collider benchmark.  For direct comparison with previous CLIC studies~\cite{Chakraborty:2018khw,Mekala:2022cmm}, we adopt identical accelerator parameters: $\sqrt{s} = 3~\text{TeV}$, $\mathcal{L}_{int} = 4~\text{ab}^{-1}$, and $(P_{e^+}, P_{e^-}) = (0, -80\%)$ beam polarization\footnote{The effects of beam polarization were not considered in the aforementioned previous analysis. The current results, which include this effect, allow for a more direct comparison with other references.}.
Our analysis significantly extends Ref.~\cite{Chakraborty:2018khw} by properly accounting for polarization effects - while the original study used unpolarized cross sections, we demonstrate that the $(0,-80\%)$ configuration enhances both signal and background processes by a consistent factor of 1.8. Using these corrected cross sections, we derive updated 95\% CL exclusion limits ($\mathcal{Z}_{\text{excl}} = 1.645$) that reflect realistic CLIC operation. Our results conclusively establish that a 3 TeV CLIC would achieve sensitivity to heavy neutrino mixing parameters that is improved by approximately two orders of magnitude relative to projected hadron collider capabilities, while maintaining competitive performance with other CLIC search channels.

\section{CONCLUSION and discussion}\label{sec:conclusion}
We present a comprehensive investigation of heavy Majorana neutrino ($N$) production and detection at a 3~TeV CLIC. In the high-mass regime (1--2.9~TeV), we analyze two distinctive decay channels where the $W$ and Higgs bosons from $N \to \ell^\pm W^\mp$ and $N \to \nu h$ decays are highly boosted, resulting in collimated hadronic decay products that form characteristic fat-jet signatures: (i) Case 1: $N \to \ell^\pm W^\mp$ with $W \to$ hadrons,  yielding the final state $1\ell + J_W + \slashed{E}_T$; and (ii) Case 2: $N \to \nu h$ with $h \to b\bar{b}$, producing $J_h + \slashed{E}_T$ final states.

 Through comprehensive detector-level simulations of both signal processes and relevant SM backgrounds, we establish the following sensitivity ranges for the heavy neutrino mixing parameter $|V_{\ell N}|^2$ in the mass range $m_N = 1000-2900~\text{GeV}$:

\begin{itemize}
    \item Case 1 ($1\ell + J_W + \slashed{E}_T$ final states):
    \begin{itemize}
        \item 1 ab$^{-1}$: $|V_{\ell N}|^2 \in [6.9\times 10^{-6}, 9\times 10^{-5}] $
        \item 3 ab$^{-1}$: $|V_{\ell N}|^2 \in [3.9\times 10^{-6}, 5.1\times 10^{-5}]$
        \item 4 ab$^{-1}$ ($-80\%$ electron polarization): $|V_{\ell N}|^2 \in [2.1\times 10^{-6}, 2.7\times 10^{-5}] $
    \end{itemize}

    \item Case 2 ($J_h + \slashed{E}_T$ final states):
    \begin{itemize}
        \item 1 ab$^{-1}$: $|V_{\ell N}|^2 \in [1.9\times 10^{-5}, 2.0\times 10^{-4}] $
        \item 3 ab$^{-1}$: $|V_{\ell N}|^2 \in [1.1\times 10^{-5}, 1.1\times 10^{-4}]$
        \item 4 ab$^{-1}$ ($-80\%$ electron polarization): $|V_{\ell N}|^2 \in [1.7\times 10^{-5}, 1.8\times 10^{-4}]$
    \end{itemize}
\end{itemize}

These findings demonstrate that the clean experimental environment of $e^+e^-$ collisions offers unique advantages for precision measurements of the neutrino mixing parameters.
While this study focuses on the heavy Majorana neutrino scenario, it is worth discussing its relation to the Dirac case. For the on-shell production considered here, the primary difference lies in the total width: a heavy Dirac neutrino has twice the width of a Majorana neutrino with the same mass and coupling, leading to a larger production cross-section and consequently stronger exclusion limits for the Dirac type under the same conditions~\cite{Mekala:2022cmm}. This suggests that our limits, when interpreted conservatively, also apply to the Dirac scenario. More importantly, future work aimed at distinguishing the two possibilities could leverage differential kinematic distributions. For instance, the lepton angular distribution in the neutrino's rest frame is expected to be forward for a Dirac neutrino but isotropic for a Majorana neutrino because of the interference between its particle and antiparticle decay modes. Methods such as analyzing the rapidity distribution peak with single lepton tagging, as proposed in~\cite{Cao:2024rzb}, may offer a clear path forward for experimental verification at future lepton colliders.

\begin{acknowledgments}
This work of Y.-B. Liu is supported by the Natural Science Foundation of Henan Province, China (Grant No. 252300421988), J.-W. Lian is supported by the National Natural Science Foundation of China (Grant No. 12447140) and the Natural Science Foundation of Henan Province, China (Grant No. 252300421771).
\end{acknowledgments}


\end{document}